# Direct Observation of Photonic Jets and Corresponding Backscattering Enhancement at Microwave frequencies

Li Zhao[*], and Chong Kim Ong

*Centre for Superconducting and Magnetic Materials, Department of Physics, National University of Singapore, 117542, Singapore,*

We experimentally verified the phenomena of photonic jets generated by plane-electromagnetic-wave-illuminated dielectric micro-cylinders with diameter comparable to the corresponding wavelength at microwave frequencies. Using a home-made 2D spatial field mapping system, we carried out a point-by-point measurement of both phase and intensity of spatial electric field distribution inside and around scattering cylinders, providing a clear complete electromagnetic field picture for these phenomena. Correspondingly, the theoretically predicted super-enhancement of the backscattering induced by small particles of deeply-subwavelength size located within the photonic jets was also confirmed. Our measurements agreed well with the numerical simulations, indicating that the photonic jets indeed can provide a promising powerful way for deeply subwavelength detection and imaging.

[*]Corresponding author: lzhao1027@ustc.edu.cn

## I. Introduction

Light scattering by dielectric 2D cylinders or 3D spheres has long been considered as a classical subject in optics [1-3]. Nevertheless, a re-investigation of this problem at a mesoscopic scale in 2004 by means of the modern high-resolution numerical simulation led to the discovery of photonic nanojets (comparable to the visible optical wavelength) [4]. Since then, there has been a growing interest in the properties of optical photonic nanojets and its possible application [5-18].

The photonic nanojet is a very narrow (with sub-wavelength waist), highly intense (several orders higher than surrounding medium) beam that emerges from the shadow side of a mesoscopic dielectric cylinder or sphere (usually only several wavelengths in diameter and with appropriate refractive indexes) illuminated by optical plane-waves. When the diameter of dielectric micro-sphere/cylinder and their optical refraction indexes (relative to background medium) are optimized, the photonic nanojets can propagate with little divergence for several wavelengths. The photonic nanojets make it possible to develop a low-cost and high-throughput method to achieve a deeply sub-wavelength focus. As for the conventional lithography techniques, the resolution is fundamentally restricted by the classical diffraction limit which is the order of the light source's wavelength. Recently a novel kind of lithography technique utilizing photonic nanojets have demonstrated that the large-sized highly uniform arrays of nanoholes or nanopillars can be easily fabricated using solely commonplace UV light source [16]. The strong localized electromagnetic fields within the photonic nanojets generated by add-in floating dielectric microspheres in solutions nanojets have been used to enhance Raman scattering [17] and the two-photon fluorescence [18].

Further numerical studies found that the presence of a deep sub-wavelength particle within a photonic jet can significantly enhance the backscattering of the micro-cylinder/sphere, in spite of the fact the diameter of the add-in small particles might be less than 1/100 of the micro-cylinder/sphere. The enhanced backscattering can be several orders of magnitude greater than that caused by the isolated nanoparticle itself. This ultra-enhanced backscattering makes it possible to develop a novel optical ultramicroscopy technique to detect and image nanoparticles such as proteins, viral particles and even single molecular, making it highly applicable in biology and chemistry. As demonstrated in recent experiments [13], they can also be used to detect deeply subwavelength pits in optical data-storage media, which enable much higher density of data storage than present techniques.

At present, to the best of the our knowledge, there are only quite a few direct visual experimental confirmations of optical photonic nanojets because of the difficulty in the fabrication of micro-sized samples with high precision and the limited resolution of present spatial detection technology in the visible light region [11]. In order to simplify the sample preparation and corresponding measurement, Heifetz et al [10] was the first to scale up the originally reported optical problem into the microwave-operating region. One apparent advantage is the much longer wavelength of microwave (for example, λ=1cm at 30GHz), which make it easy to fabricate virtually perfect samples by machining alone. Additionally, the high-precision microwave detection is readily available nowadays. Heifetz et al used a vector network analyzer (VNA) to generate the 30GHz electromagnetic wave to illuminate a acrylic sphere(with 7.62cm in diameter and permittivity $\varepsilon = 2.57+0.008j$), and measured the reflection signal. When the small metal particles were introduced in the jet region, they observed the greatly enhanced backscattering, which agreed well with numerical simulation by finite difference time domain (FDTD) method. Later, similar experiments by Kong et al demonstrated the robust detection of deep sub-wavelength pits using this enhanced backscattering effect also at 30GHz [13].

However, the above-mentioned two microwave experiments measured only the backscattering intensity at the port of excitation source. A physical picture comprehensively describing the whole structure of spatial electromagnetic fields is still lacking. In 2006, Smith et al in Duke University develops a point-by-point phase-sensitive method of field measurements to obtain the spatial field mapping of metamaterials samples loaded in a two dimensional (2D) parallel plate waveguide [19]. Using this field mapping apparatus, they performed the state-of-art demonstration of metamaterial electromagnetic cloak at microwave frequencies(X-band) [20]. Inspired by their work, we carried out a visualization of the photonic phenomena at microwave region(X band) for the first time, using a analogous home-made spatial field mapping system. Correspondingly, the enhanced effect of backscattering was also confirmed and quantitatively characterized. All these results agreed well with our numerical simulation by finite element method.

The details about our experimental setup have been described in the following Section II. In Section III, we present the results and discussions on field mapping of photonic jets, and corresponding measurements of backscattering enhancement, as well as the comparison with the numerical simulation. The conclusion is presented in Section IV

## II.  Experimental Method and Materials

### 1.  Experimental Setup

The whole setup of our field mapping system is schematically shown in Fig. 1(a). The basic geometry of the system is a parallel plate waveguide composed of two aluminum plates, separated by an 11mm air-filled spacing. The larger fixed top aluminum plate is 1.2m x 0.6m x 5mm, and supported by four pillars whose height can be adjustable respectively. The smaller bottom plate is 0.6m x 0.4m x 3 mm, and mounted on a movable XY-stage. This stage is driven by two PC-controlled step motors, which allow it to move freely in two dimensions relative to the fixed top metallic plate. The travel range in each direction (X and Y) is more than 300mm and the stage has a resolution of 2D positioning (<50μm) and good repeatability.

An X-band (8.2-12.4GHz) waveguide adapter is fixed to one edge of the bottom plate as a source feed. The microwave electromagnetic fields introduced by the adapter are restricted between the two metallic plates. To probe the field distribution into the chamber, we use a coaxial detection antenna inserted into the waveguide chamber through a small hole (with the diameter of 2mm) in the center of the top plate. The antenna does not protrude below the top plate in order to reduce unwanted disturbance to the intrinsic field distribution in the chamber. Its central probe pin was set just flush with the lower surface of the fixed top plate.

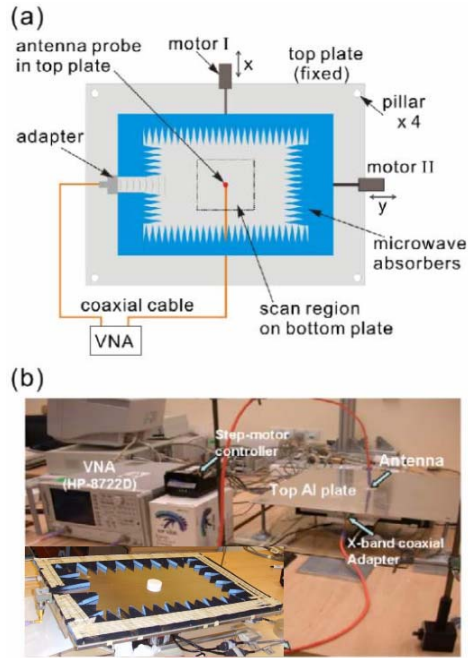

Fig. 1. (Color online) Schematic top view (a) and photograph (b) of our 2D spatial field mapping system. The inset in (b) is the photograph of the sampled-loaded chamber, with the top plate removed.

For air-filled parallel plate waveguides, the basic TEM mode has no cut-off frequencies while the other modes ($TE_n$ or $TM_n$, integer $n \geq 1$), have the respective cut-off frequencies, $f_c = c/(2nd)$ ($c$ is the velocity of light in the air, and $d$ the spacing of two conducting planes). Therefore, it is possible to excite only TEM mode inside the chamber, in which the electric and magnetic fields do not vary along the axis normal to the conducting planes (z-axis). The z-axis translation invariability holds unless the inserted scattering objects between the two plates break this symmetry. In our field mapping apparatus, the spacing between the two parallel plates is about 11mm. The X-band feed source doesn't excite any TE or TM modes because the lowest cut-off frequency $f_c$ of the TE or TM mode (n=1) is about 13.6GHz for d=11mm.

For microwave measurement, we use a highly sensitive vector network analyzer (VNA) (HP8722D, Agilent Technologies). Two flexible low-loss coaxial cables were used to connect the two VNA ports to the adapter and antenna respectively. The VNA provides the microwave source excitation to the adapter feed from port 1 and detects the return signal at port 2 from the detection antenna. In another word, the transmitted parameter $S_{21}$ (the complex ratio of test signal to the excitation source) is measured at different positions of the antenna relative to the waveguide chamber. Because of the z-axis translational invariance, the electromagnetic field obeyed the 2D scalar wave equation. The measured $S_{21}$ actually reflects the local complex electric field directly, i.e.

$S_{21} \propto E(x,y) = |E(x,y)|e^{j\theta} = \text{Re}[E(x,y)] + j\text{Im}[E(x,y)]$.

The scattering samples are placed on the center of bottom plate. They are nearly of the same height (10 mm) as the plate separation. By stepping the lower plate in small increments and recording both amplitude and phase of $S_{21}$ parameter at each position, a full 2D spatial distribution of the microwave scattering field inside the chamber can be mapped intactly. We use 11mm-thick saw-tooth-like microwave absorbing foams to enclose the mapping range and leave an inlet for the incident microwave from the adapter feed. The absorbing foams can eliminate the back reflection from the boundaries of the chamber and suppress the possible resonant modes in the chamber.

Fig. 1(b) is the photograph of our integrated field mapping system. The bottom-left inset is the photograph of the sampled-loaded chamber with top plate removed.

To test the performance of our system, the empty chamber without any scatters has been tested and quantitatively analysis has been presented in [21]. A quite good signal-to-noise level with highly repeatability was achieved. There is no discernable reflection from the absorbing boundaries, except a slightly decaying background. The decaying in magnitude comes mainly from the transverse expansion of the incident quasi-cylindrical wave during its propagation, and approximately obeys the $r^{-1}$ law ($r$ is the distance from the small aperture of the feed source port). So the decaying background can be easily subtracted from the raw data in the analysis process in further experiments. Using this field mapping system, we have successfully characterized different microwave absorbing materials superior to traditional testing methods. Further details about our experimental apparatus and corresponding data processing can be found in [21].

## 2. Materials

The targets of illumination to generate photonic jets in our experiments are cylinders made of Teflon (from Lucite International) by mechanical processing, with diameter of 60mm and height of 10mm. The processed surfaces were very smooth. Their dielectric properties have been measured by the transmission line method [22] using an X-band wave-guide (WR90) and the HP 85071(a commercial materials measurement software module). The permittivity of our Teflon samples over the entire X-band is almost constant, $\varepsilon = 2.10+0.001j$.

## III. Results and discussions

### 1. Direct Imaging of Photonic Jets

Firstly, we placed a Teflon cylinder with 60mm in diameter in the central region of the chamber. We pre-scanned an area of 300mm by 210mm, almost to the utmost limit of the present measuring system. The mapping of $S_{21}$ (at X band) was scanned in sparse steps of 3mm to get a quick full view of the whole field distribution in the chamber. The typical spatial field maps of the electric field intensity (square value of the magnitude, corresponding to $|S_{21}|^2$) and real part of the electric field (corresponding to the real part of $S_{21}$, Re[$S_{21}$]) at 10 GHz are depicted in Fig. 2(a) and (b) respectively.

The incident microwave field propagates in the positive x-axis direction. The curvature of the wave fronts, i.e. bright and dark "ripples" shown in Fig. 2(a), indicates that the incident microwave is quasi-cylindrical-waves-like, which results from the small aperture of the feed (a X-band waveguide adapter, about 2.3cm x 1.0cm) and the finite distance from the adapter to the front side of the scanning region (about 200mm). Nevertheless, the central area region can be treated as a good approximation to the case as illuminated by the plane wave in free space. Furthermore, the radial space between the adjacent wave fronts (in air-filled region) is 30mm. As we expected, it equals the exactly wavelength of the electromagnetic wave at 10GHz.

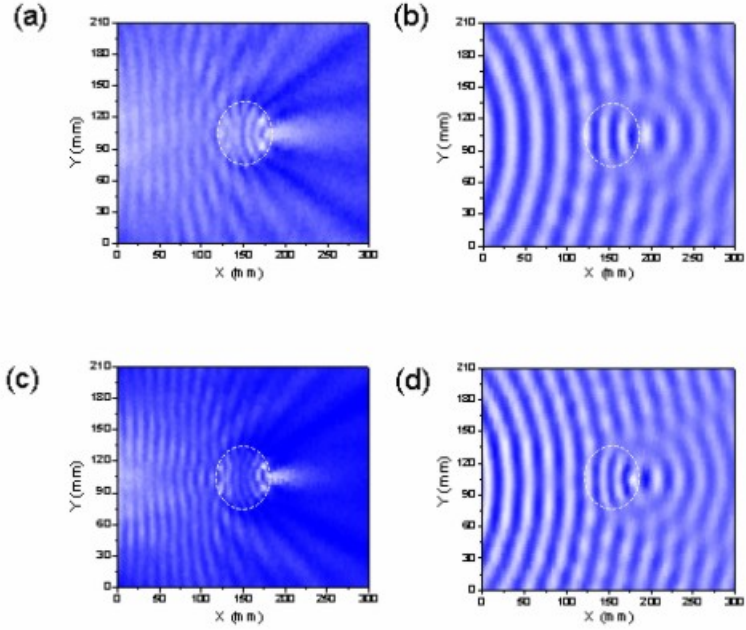

Fig. 2. (Color online) Spatial distribution of the intensity (a) and real part (b) of the electric field measured at 10GHz in in the chamber loaded with a Teflon cylinder (60mm in diameter, marked as enclosed spacing by the dashed line). The 300mm x 210mm mapping region was scanned in steps of 3 mm.(c) and (d) are corresponding results measured at 12GHz.

In the corresponding map of total electric field intensity, $|S_{21}|^2$, shown in Fig. 2(b), we can see the modulated pattern in the front of the dielectric cylinder, which comes from the interference effect of the incident waves and scattered waves by the cylinder. The most noteworthy phenomena appear in the geometrically optical shadow region. A narrow, bright beam generates from the cylinder's surface on the shadow side. It propagates a distance nearly 2λ (λ=30mm for 10GHz) with low divergence.

In Fig. 2 (c) and (d), the field map over the same scanning region but acquired at different frequency (12GHz) is presented. The phenomena of photonic jet still exist with slightly narrower waist. As the frequency of incident electromagnetic wave varies, similar results were found over the whole X band in our experiments. The robustness of photonic jets rules out the origin of dielectric resonance, which makes photonic nanojets more suited for potential brand-band applications.

In order to obtain a more detailed description about the photonic jet, we rescan a sub-region (150mm by 90mm) including photonic jet in much finer steps (only 1mm) to achieve a higher spatial resolution. A typical intensity map of field at 12GHz is shown in Fig. 3(a). For comparison, we also carried out a full wave simulation using the COMSOL Multiphysics finite-element–based commercial software package. For simplicity, we only considered the ideal case in which the cylinder was illuminated by plane wave. The perfect magnetic conducting (PMC) boundaries are used on two lateral sides of computational domain, which constrain the polarization of the incident TEM wave. At the end side of region, we employed the perfectly matched layer (PML) as the boundary to simulate the microwave absorber. The calculated distribution of field intensity is presented in Fig. 3(b). Qualitatively, the agreement between the simulation and experiment is quite good, considering the real-world effects such as non-ideal absorbing boundaries and quasi-cylindrical incident waves under experimental conditions.

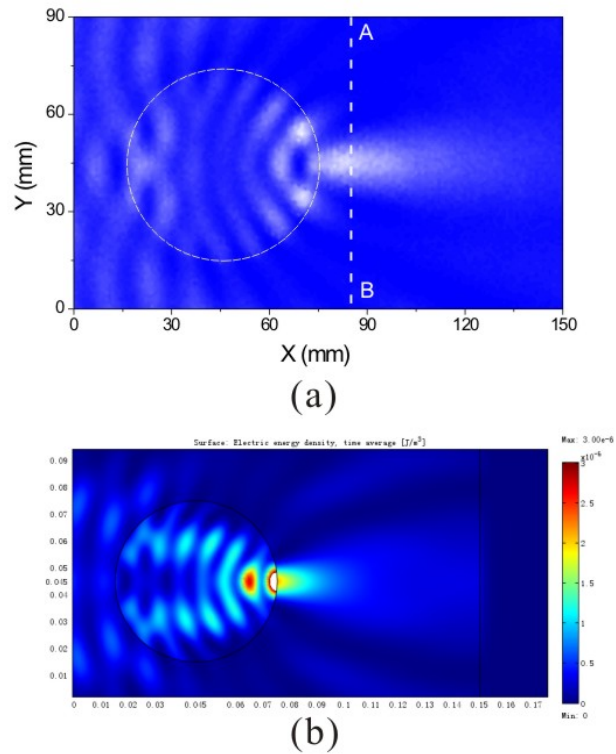

Fig. 3. (Color online) Spatial distribution of the field intensity at 12GHz(a) in the chamber loaded with a 60mm-diameter Teflon cylinder. The 150mm x 90mm mapping region was scanned in steps of 1 mm. (b) is the corresponding numerical simulation using COMSOL.

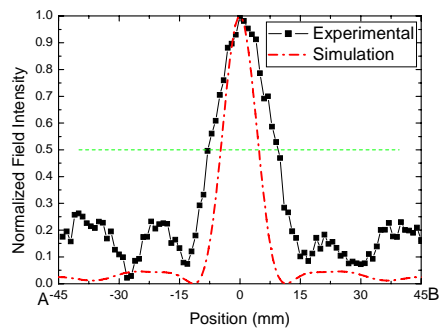

Fig. 4. (Color online) Normalized field intensity cross section along the AB line(shown in Fig. 3(a)) at12GHz. The dashed line is the corresponding results from the simulation using COMSOL.

For further quantitative study of the sub-wavelength waist of photonic jets, we plotted the line-scans of the field intensity along the transversal AB line (shown in Fig. 3(a), which is about 10mm away from the end surface of the Teflon cylinder. Both experiment and simulation data are shown in Fig 4. The measured FWHM (full width at half maximum) of the waist of photonic jets is only about 0.65λ (λ=25mm for 12GHz). Though still broader than the numerical simulation (about 0.4λ), it is much less than the diffraction limit of the conventional optical focus.

## 2. Enhanced Backscattering

In order to perform the backscattering measurements, we removed the antenna probe fixed in the top plate of our apparatus. Instead, a straight thin copper wire (2mm in diameter) was vertically inserted into the chamber, through the probe hole in the top plates. The lower end of the wire nearly touches the surface of the bottom and the z-axis translation invariability holds in this configuration. This wire serves as a deeply sub-wavelength scatter in our experiments. Its diameter is only 1/15λ at 10GHz (λ=30mm), corresponding a 30nm nanoparticles for visible light with λ=450nm.

The backscattering is probed by measuring reflective $S_{11}$ parameter from the feed adapter connected to port 1 of the VNA. To null out the background reflection signals mainly from the impedance mismatch at the aperture of feed adapter and the surface of the dielectric cylinder, firstly we measure the $S_{11}$ parameter for the chamber without inserted wire except the Teflon cylinder, which is used as background reference level, $S_{11}^0$. Subsequently, the copper wire is inserted and the position-dependent backscattering measurement is performed as schematically shown in Fig. 5(a). The copper wire moves step-by-step along the AB path marked in fig. 5(a) (the same line as shown in Fig. 3(a)), across the photonic jet region. The acquired reflective signal $S_{11}^1$ at each step was normalized by the background reference level to give the backscattered intensity perturbation by the inserted metal wire, $\left|S_{11}^R\right| = \left|S_{11}^1 / S_{11}^0\right|$, which is the same definition as used by Heifetz et al in [10].

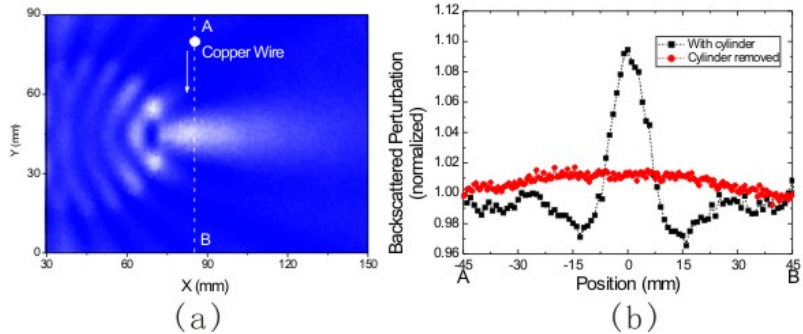

Fig 5. (a) Scheme about a copper wire with diameter of 2mm moving through a photonic jet . (b) Black line: The measured normalized backscattered perturbation, $\left|S_{11}^R\right|$, as a function of the wire position at 10GHz; red line: corresponding results when the Teflon cylinder is removed.

The strong position-sensitive backscattered perturbation, $|S_{11}^R|$, generated by the thin copper wire is shown in Fig. 5(b). As entering the subwavelength jet region of high intensity, the deeply subwavelength metallic object can greatly perturb the highly localized field and therefore leads to super-enhanced backscattering, which forms a sharp contrast to the non-jet region. In addition, the thin wire also causes apparent perturbation as enter into the two side lobes, which are also generated by the cylinder and adjacent to the center jet as shown in Fig. 3. For further comparison, we removed the Teflon cylinder and measured the backscattering caused by the metallic wire alone (red lines shown Fig. 5(b). As expected, the backscattering by the metallic wire alone is very small and no apparent position-dependent structure can be discerned for the wire's cross section is much smaller than the wavelength.

As shown in Fig 5(b), the FWHM of the position-dependent scattering peak is only 0.5λ. Therefore, this high position-sensitive feature of backscattering enhancement can be utilized to develop a promising technique for accurate positioning and even manipulating nanoparticles using visible light.

The inserted object in the jet region doesn't always enhance the backscattering. The strongly position-dependent interaction in jet region may enhance or depress the backscattering. In fact, as suggested by Heifetz et al, along the central line of photonic jet, $|S_{11}^R|$ oscillates as the distance of the inserted particle from the top surface varies shown in Fig. 3 in [10]. To get a comprehensive picture about the dependence of the backscattered perturbation on the location of the copper wire, we scanned the wire location in the region behind the Teflon cylinder (i.e. 75-150mm along X axis and 0-90mm along Y one in Fig. 5(a)) with the 1mm-step. The results of $|S_{11}^R|$ is 3D spatially plotted in Fig. 6.

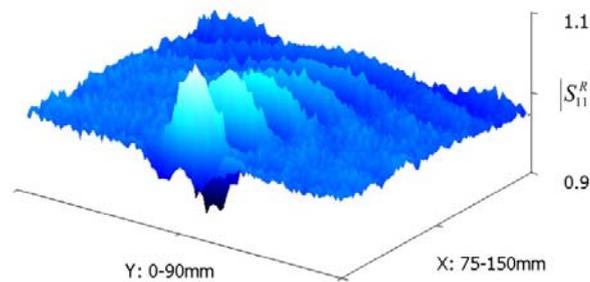

Fig 6. Measured normalized backscattered perturbation, $|S_{11}^R|$, as a function of the planar position of the inserted copper wire at 10GHz.

From Fig. 6, we see that $|S_{11}^R|$ oscillates along the center line of the photonic jet. This behavior coincides quite well with Heifetz's experimental results, as well as FDTD calculations [10]. The strongest deviation from the non-jet background region occurs approximately in the spot of the jet region. This high sensitivity of photonic-jet-enhanced backscattering is extremely useful for accurate resolution of the location of subwavelength insertion, which can be used in the spatial detection and trapping of the nanoparticles using visible light in future.

## IV. Summary

Using our home-made 2D spatial field mapping measure system, we obtained a visual confirmation of the photonic jets at microwave frequencies (X-band) for the first time. The measurements agreed well with corresponding numerical simulations. We also observed the associated enhanced backscattering effect induced by scatters of deeply subwavelength size located in the photonic jet region. This method offer a powerful experimental way to 'scale up' micro-optical problems to microwave region and can provide a comprehensive electromagnetic field picture for them. Our present results in this paper illustrates that the photonic jets can be utilized to develop a promising powerful technique for subwavelength detection, imaging and even manipulating in the future

## Acknowledgement

We would like to thank Dr Y. G. Ma and P. Wang for our interesting discussion as well as Mr. X. Chen and S. Sheng for their technical assistance. The project was supported by the Defense Science and Technology Agency in Singapore.